\pdfoutput=1

\documentclass[11pt]{article}

\usepackage[]{ACL2023}

\usepackage{times}
\usepackage{latexsym}

\usepackage[T1]{fontenc}

\usepackage[utf8]{inputenc}

\usepackage{microtype}

\usepackage{inconsolata}

\usepackage{graphicx}
\usepackage{subcaption}
\usepackage[normalem]{ulem}

\usepackage{tikz}
\usetikzlibrary{arrows}
\usepackage{tikz}

\usepackage{tikzscale}
\usetikzlibrary{arrows.meta,
                calc,
                positioning,
                quotes,
                shapes.geometric}

\pgfdeclareshape{datastore}{
  \inheritsavedanchors[from=rectangle]
  \inheritanchorborder[from=rectangle]
  \inheritanchor[from=rectangle]{center}
  \inheritanchor[from=rectangle]{base}
  \inheritanchor[from=rectangle]{north}
  \inheritanchor[from=rectangle]{north east}
  \inheritanchor[from=rectangle]{east}
  \inheritanchor[from=rectangle]{south east}
  \inheritanchor[from=rectangle]{south}
  \inheritanchor[from=rectangle]{south west}
  \inheritanchor[from=rectangle]{west}
  \inheritanchor[from=rectangle]{north west}
  \backgroundpath{
    \southwest \pgf@xa=\pgf@x \pgf@ya=\pgf@y
    \northeast \pgf@xb=\pgf@x \pgf@yb=\pgf@y
    \pgfpathmoveto{\pgfpoint{\pgf@xa}{\pgf@ya}}
    \pgfpathlineto{\pgfpoint{\pgf@xb}{\pgf@ya}}
    \pgfpathmoveto{\pgfpoint{\pgf@xa}{\pgf@yb}}
    \pgfpathlineto{\pgfpoint{\pgf@xb}{\pgf@yb}}
 }
}

\usepackage{adjustbox}

\usepackage{hyperref}

\title{Characterizing Financial Market Coverage using Artificial Intelligence}

\author{Jean Marie Tshimula,$^{*1,2}$ D'Jeff K. Nkashama,$^{*1}$ Patrick Owusu,$^{*1,3}$ Marc Frappier,$^{1}$ \\ {\bf Pierre-Martin Tardif,$^{1}$ Froduald Kabanza,$^{1}$ Armelle Brun,$^{3}$ Jean-Marc Patenaude,$^{4}$}  \\ {\bf Shengrui Wang,$^{1}$  Belkacem Chikhaoui $^{2}$} \\
  $^{1}$Department of Computer Science, Université de Sherbrooke, QC J1K 2R1, Canada \\
  $^{2}$LICEF Research Center, Université TÉLUQ, QC H2S 3L5, Canada\\
  $^{3}$LORIA, Université de Lorraine, 54000 Nancy, France \\
  $^{4}$Laplace Insights, QC J1H 1P9, Canada \\
  {\tt shengrui.wang@usherbrooke.ca} \\ }

\begin{document}

\maketitle

\def\thefootnote{*}\footnotetext{These authors contributed equally to this work}\def\thefootnote{\arabic{footnote}}

\begin{abstract}

This paper scrutinizes a database of over 4900 YouTube videos to characterize financial market coverage. Financial market coverage generates a large number of videos. Therefore, watching these videos to derive actionable insights could be challenging and complex. In this paper, we leverage Whisper, a speech-to-text model from OpenAI, to generate a text corpus of market coverage videos from Bloomberg and Yahoo Finance.  We employ natural language processing to extract insights regarding language use from the market coverage. Moreover, we examine the prominent presence of trending topics and their evolution over time, and the impacts that some individuals and organizations have on the financial market. Our characterization highlights the dynamics of the financial market coverage and provides valuable insights reflecting broad discussions regarding recent financial events and the world economy. 

\end{abstract}

\section{Introduction}

Financial markets, especially the stock market, enjoy substantial coverage day-to-day on digital platforms such as YouTube. Besides the presenters, experts with much understanding of the stock markets work as contributors or panelists who share their perspectives on various topics. On YouTube, channels such as Yahoo Finance's Stock Market Coverage provide a wealth of information about the development of financial market events, that can allow the audience to get informed on trending topics, among others. Most studies on the impact of stock news, for instance, on stock prices focused on using headlines from renowned news agencies or blog posts \cite{Jariwala:20,VelayDaniel:18,NemesKiss:21}.
The news coverage particularly gives a topic context and meaning \cite{Chipidza:21}, much as financial television (TV) programs such as CNBC Markets.
However, in comparison to traditional news coverage, the dissemination of financial and economic news is either a segmented section or a dedicated channel on TV.

While most financial market studies focused on data sources such as news headlines, financial reports, and social media posts, continued news coverage of any kind have had limited usage for the purpose of analysis. 
In particular, one may argue the authenticity of social media posts from either Facebook or Twitter, as misinformation is ubiquitous on such platforms \cite{Kogan:18}. With media coverage, not only is it factually oriented to decrypt market news and events, an advantage is that the information is fact-checked. Besides this, the viewpoints of renowned experts can get contradicted, backed, or completed by journalists or other high-profile persons in the world of finance and economics.

Given the popularity of YouTube, the publicly available videos constitute a reliable source of data for further analysis. However, the challenge of analyzing videos, in general, requires either capturing image frames or manipulating snippets of an entire video \cite{Snelson:21}.
The social science field is an example where videos are a data source for analysis; these are either non-verbal or otherwise \cite{Luff:12}. 

In this paper, we build corpora of transcribed videos on YouTube that focus on the financial market and the economy. We used OpenAI's Whisper \cite{Radford:22} to transcribe videos to texts since market coverage generates a large volume of video data. Consequently, watching tons of financial market coverage videos to derive actionable insights can be challenging and complex. To this end, we transcribe market coverage videos to texts for simplifying analyses. Specifically, we use a topic modeling approach for generating topics related to the markets. Further, we perform an n-gram analysis to understand the coverage narratives and extract the most frequently mentioned persons and organizations in the market coverage using named entity recognition. It should be noted that we kept topics related to the economy and markets. 

\noindent
\textbf{Background and Related Works.}  The characterization of financial and economic news has been explored in numerous aspects, including emotions and sentiments \cite{Schumaker:12,Griffith:20} from social media posts to news headlines \cite{Mitra:11,Bukovina:16}. The role of news has been covered in \cite{Baker:21} where the authors pose the question ``{\it What drives big moves in national stock markets?}''. According to the study, news about US economic and policy developments significantly impacts worldwide equity markets. Considering previous works, most have centered on how media coverage affects financial markets \cite{Fang:09,Dogal:12,Strycharz:18}. In comparison with previous works on the effects of media coverage on financial markets, we focus on how financial and economic news coverage narrations between multiple platforms are similar. Our work is closely related to the studies of \cite{Piao:15,McBeth:18,Bhargava:22}, who compare the narratives of news coverage. Relative to these works, we investigate the evolution of the popular topics addressed over time by financial media channels on YouTube and discover similarities between the topics addressed by different media channels. Additionally, we show that financial news coverage is centered on organizations and individuals: where individuals are either heads of organizations or knowledgeable panelists or experts in finance and economics. We focus on the following two research questions. \textit{RQ1:} How are major financial events identified through language use within news coverage topics? \textit{RQ2:} To what extent do news coverage topics exhibit content coordination regarding major financial events and entities (such as organizations and individuals) across different news channels?

Specifically, this work makes the following contributions. (1) We show how effective our data collection and pre-processing strategy is for gathering digital videos and generating text information, which relates to the financial market and economic discourse. (2) We compare the narratives between reliable media channels; one of the advantages of utilizing datasets from these media channels is that they do not stem from bots. (3) We investigate the evolution of the topics addressed over time and examine the most frequently mentioned entities (organizations and persons) in financial market coverage and discover similarities between topics. (4) We publicly release our code as open source to support continued development.\footnote{\url{https://github.com/djeffkanda/market_coverage_analysis}}

\begin{figure*}[t!]
  \begin{subfigure}[b]{0.5\linewidth}
    \centering
    \includegraphics[scale=0.46]{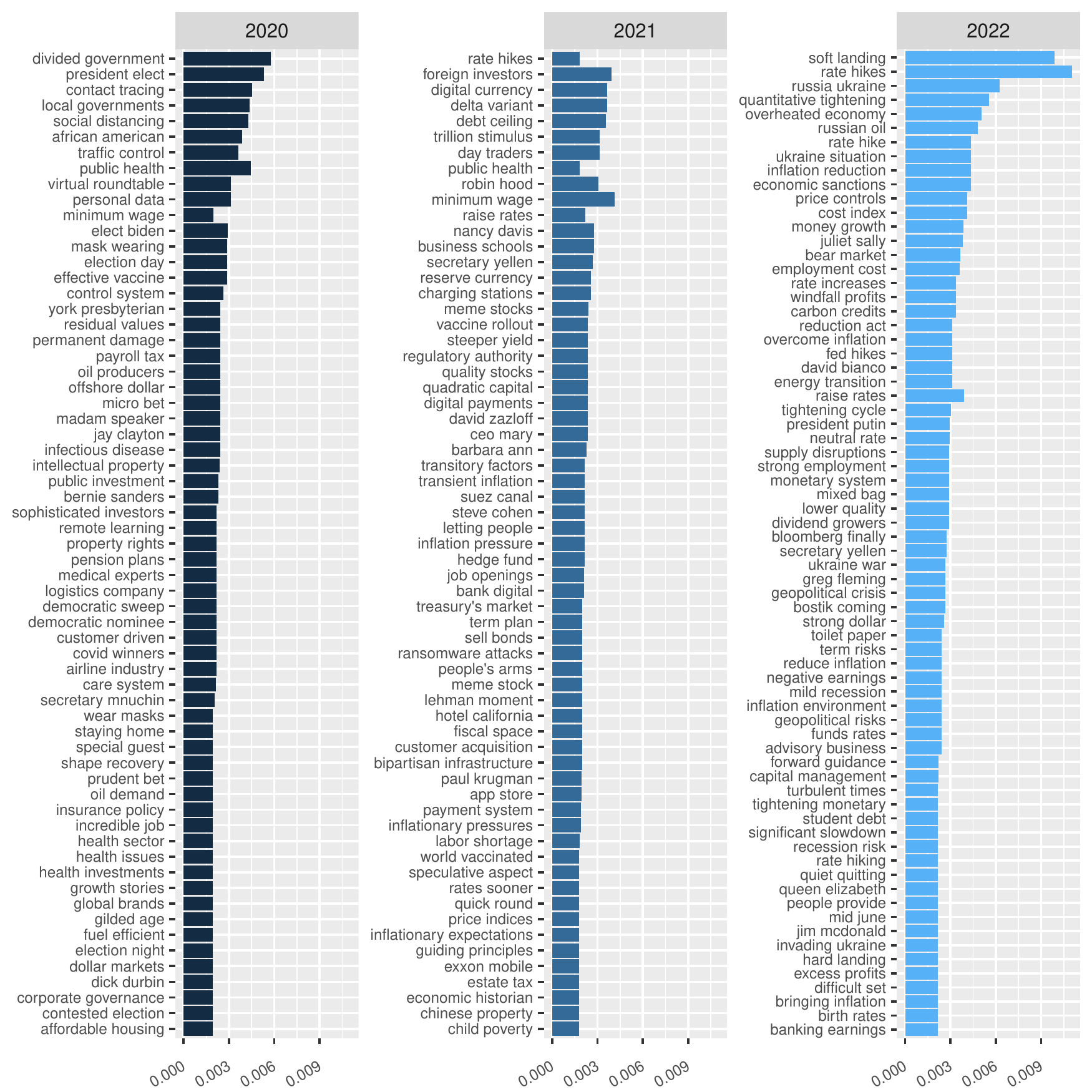} 
    \caption{Bloomberg Wall Street Week (BLW)} 
    \label{fig:coverage_bws_ngrams} 
  \end{subfigure}
  \begin{subfigure}[b]{0.5\linewidth}
    \centering
    \includegraphics[scale=0.46]{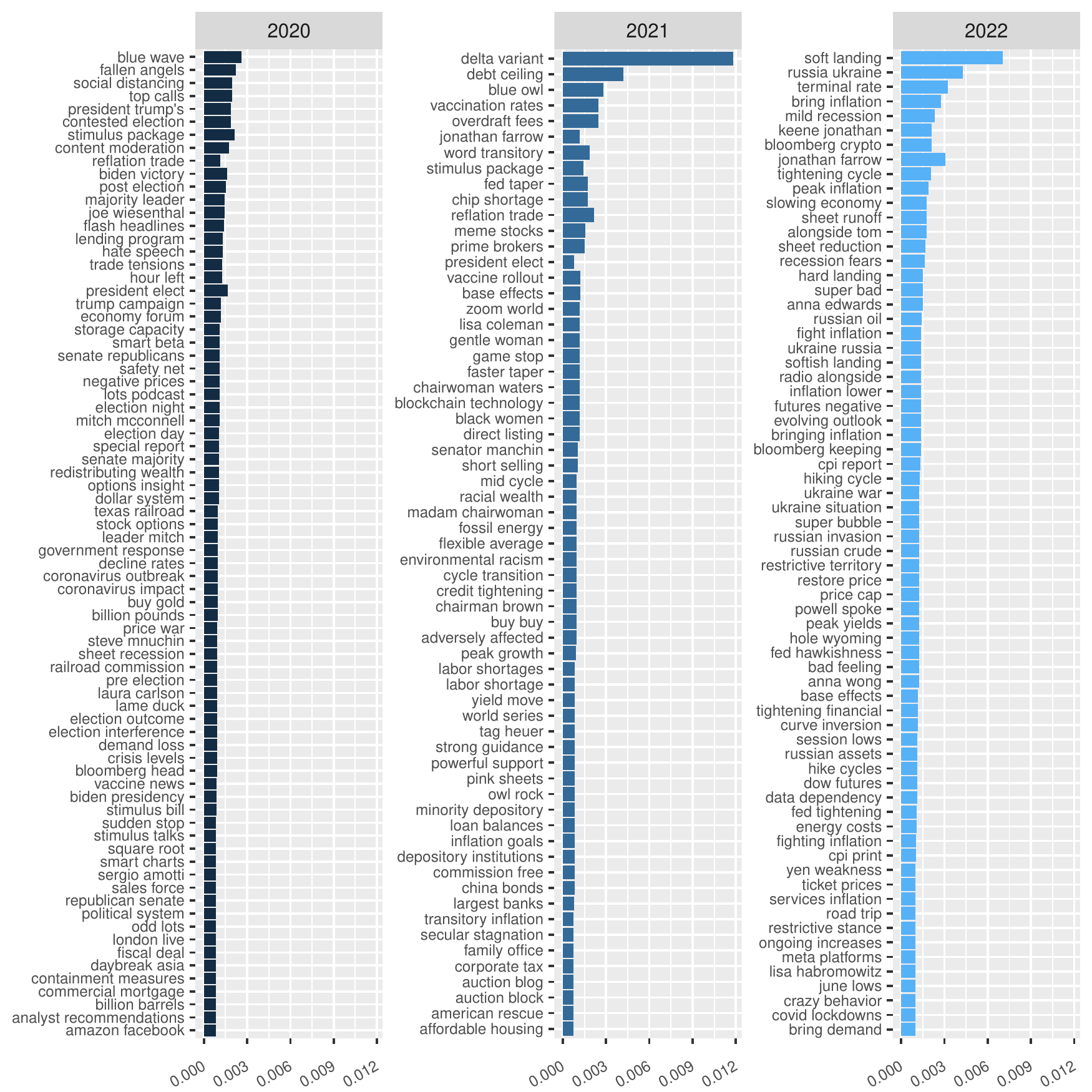} 
    \caption{Bloomberg Stock Market News and Analysis (BSM)}
    \label{fig:coverage_bsm_ngrams} 
  \end{subfigure}
  \caption{Bi-grams with the highest tf-idf from Bloomberg data. Note that \textit{x-axis} represents tf-idf values.}
  \vspace{-5mm}
\end{figure*}

\begin{figure}[htb]
    \centering
    \includegraphics[scale=0.45]{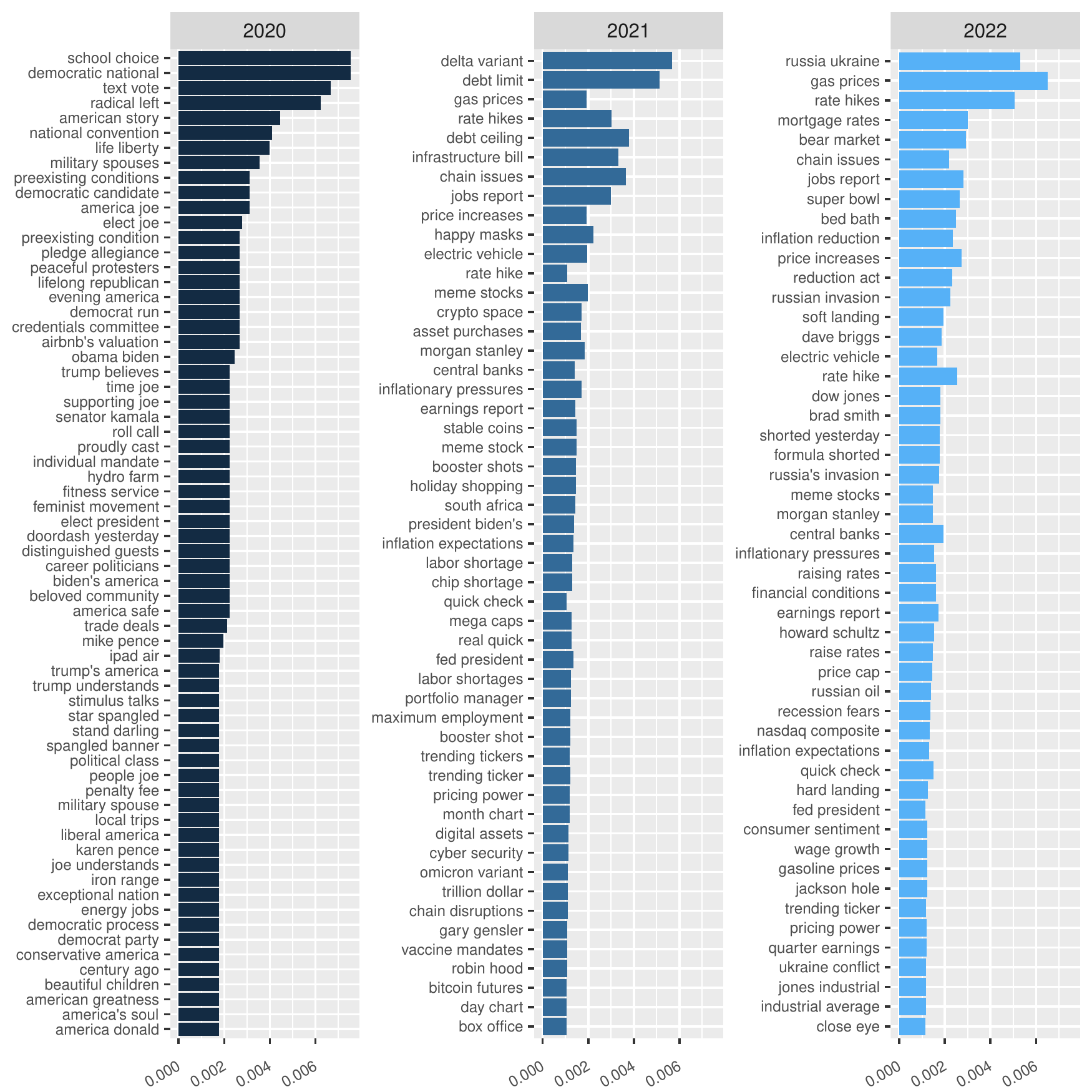}
    \caption{Bi-grams with the highest tf-idf from YFM. Note that \textit{x-axis} represents tf-idf values.}
    \label{fig:coverage_yfm_ngrams}
    \vspace{-5mm}
\end{figure}

\begin{figure*}[t!]
  \begin{subfigure}[b]{0.5\linewidth}
    \centering
    \includegraphics[scale=0.46]{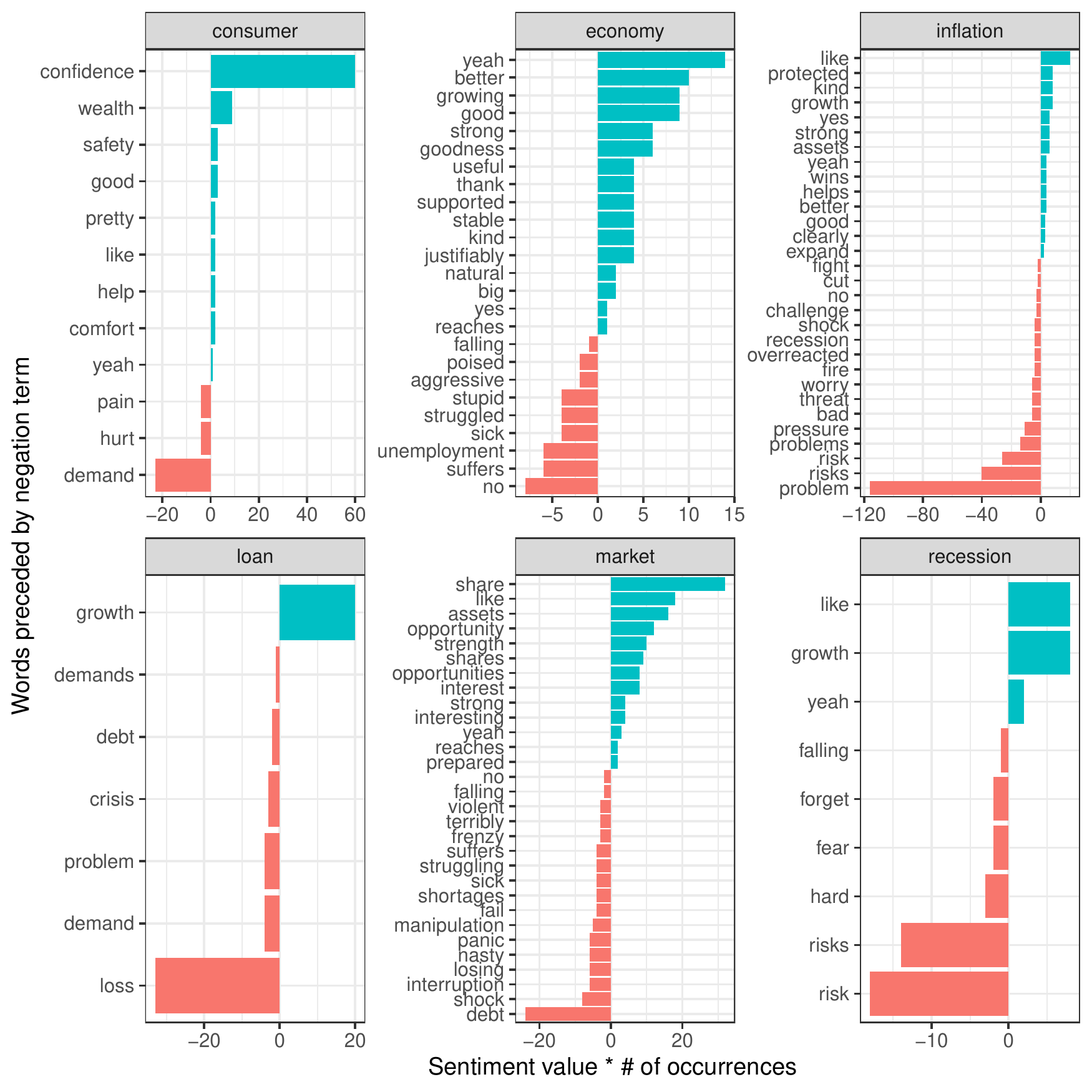} 
    \caption{Bloomberg Wall Street Week (BLW)} 
    \label{fig:sentiment_bws} 
  \end{subfigure}
  \begin{subfigure}[b]{0.5\linewidth}
    \centering
    \includegraphics[scale=0.46]{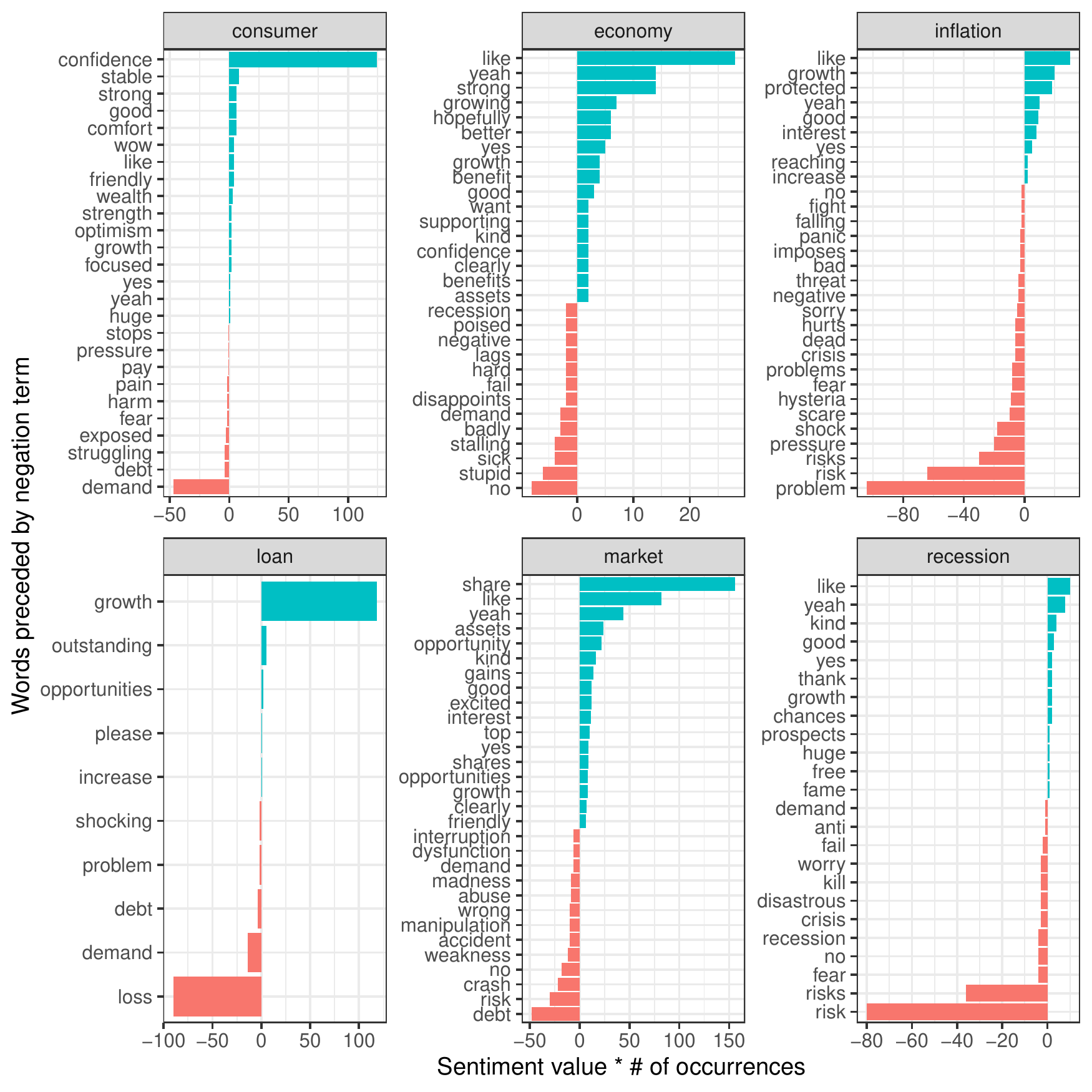} 
    \caption{Bloomberg Stock Market News and Analysis (BSM)}
    \label{fig:sentiment_bsm} 
  \end{subfigure}
  \caption{Words preceded by either \textit{consumer, economy, inflation, loan, market} or \textit{recession} that had the greatest contribution to sentiment values, in a positive or negative direction in Bloomberg coverage.}
  \vspace{-5mm}
\end{figure*}

\begin{figure}[htb!]
    \centering
    \includegraphics[scale=0.45]{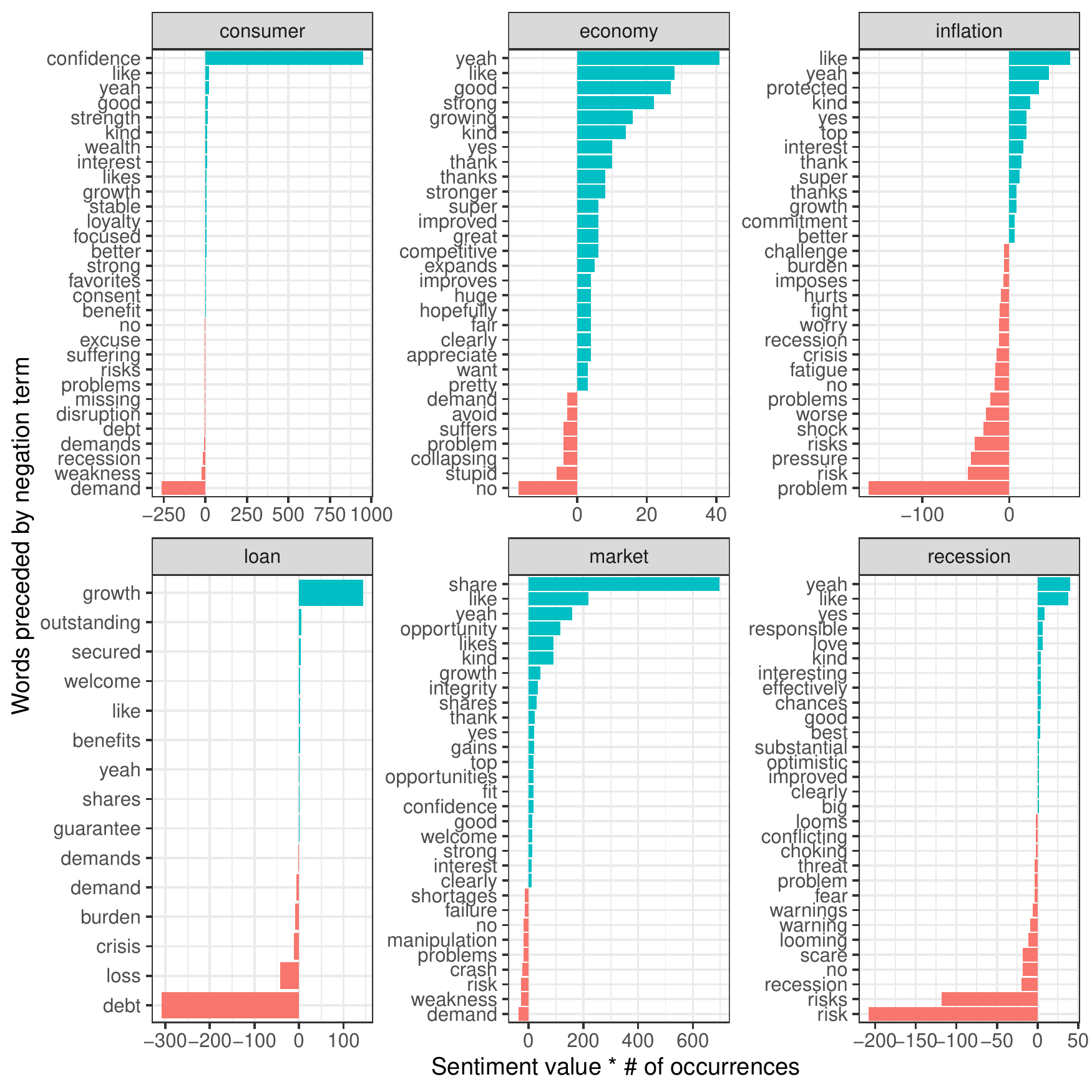}
    \caption{Words preceded by either \textit{consumer, economy, inflation, loan, market} or \textit{recession} that had the greatest contribution to sentiment values, in a positive or negative direction in YFM.}
    \label{fig:sentiment_yfm}
    \vspace{-5mm}
\end{figure}

\section{Methods}

\subsection{Data collection}

The data we used in the study was collected from the YouTube channels of Yahoo Finance and Bloomberg Markets and Finance and transcribed using the OpenAI's Whisper speech-to-text model described in \cite{Radford:22}. Note that our choice of using Yahoo Finance and Bloomberg is motivated by the fact that they (i) are among the world leaders in business news and real-time financial market coverage, (ii)  provide financial news, data, and commentary including stock quotes, press releases, financial reports, original content, and video to the world of finance every Monday–Friday from 9 am to 5 pm (ET), and (iii) host world-class specialists to express their opinions and discuss market news and events; additionally, they are freely accessible on YouTube. They decompose the markets and real-life financial issues for individual investors, industry leaders, and those seeking to invest in their future. 

\begin{table}
\renewcommand{\arraystretch}{1.5}
\setlength\tabcolsep{3.7pt}
\centering
\fontsize{8.5pt}{8.5pt}\selectfont
\caption{Data summary of the collected coverage}\label{TblOne45XTiOH2vPq1}
\begin{tabular}{ c c c c c}
\hline 
Media & Tot. collected files & Total hours & Avg. time per file \\
\hline
{BLW} & {744} & {171.16} & 14~minutes  \\
{BSM} & {3885} & {398.15}  & 6~minutes  \\
{YFM} & {318} & {2467} & 8~hours \\
\hline
\end{tabular}
\vspace{-5mm}
\end{table}

Specifically, we collected one of the YouTube playlists of the Yahoo Finance Market's official channel called Stock Market Coverage, from 02 January 2020 to 30 September 2022. We extracted two different YouTube playlists from Bloomberg and Finance's official channel, namely Wall Street Week, and Stock Market News and Analysis, for the same time period.

{\bf Yahoo Finance Stock Market Coverage (YFM)} receives top names in finance and economics to scrutinize the latest market news and contribute with cogent evidence to explain the development of the market events, identify any untapped needs in the marketplace, and provide advanced analyses and opinions.

{\bf Bloomberg Wall Street Week (BLW)} hosts influential personalities in finance and economics to talk about the week's biggest issues on Wall Street.

{\bf Bloomberg Stock Market News and Analysis (BSM)} is a playlist of Bloomberg Markets and Finance's YouTube official channel where experts discuss the latest market news and effectuate market analysis in real-time coverage.

Table \ref{TblOne45XTiOH2vPq1} summarizes statistics of the collected market coverage. For the three targeted YouTube playlists, BLW, BSM, and YFM, respectively, we extracted 744, 3885, and 318 videos for which the total hours approximate 171.16, 398.15, and 2467 and the average duration of videos counts 14 minutes, 6 minutes and 8 hours.

We utilized the OpenAI's Whisper, a speech recognition model, to transcribe audios of the collected data to text corpora \cite{Radford:22}. Speech recognition remains a challenging problem in artificial intelligence and machine learning \cite{Chiu:18,Qin:19,YZhang:20}. In a step toward solving it, OpenAI introduced Whisper, an automatic speech recognition system that approaches human-level robustness and accuracy in English speech recognition. Whisper outperformed the state-of-the-art speech recognition systems by leaps and bounds and has received immense interest for its multilingual transcription and translation capabilities spanning nearly 100 languages. Whisper was trained on 680,000 hours of multilingual and `multitask' data collected from the web, which lead to improved recognition of unique accents, background noise, and technical jargon. One of the advantages of Whisper is that it performs well even on diverse accents and technical language and is almost human-level in terms of recognizing speech even in extremely noisy situations. The architecture and the performance of Whisper over other speech recognition systems are briefly explained in its original paper \cite{Radford:22}.

Specifically, we utilized the transcribed corpora (Table \ref{TblOne45XTiOH2vPq1}) to extract insights using natural language processing techniques including n-gram analysis (\S\ref{ngram}), topic modeling (\S\ref{topicmodels}) and named entity recognition (\S\ref{ner}). We removed stopwords, numbers and special characters for performing n-gram analysis \S\ref{ngram} and opic modeling \S\ref{topicmodels}. 

\subsection{N-gram analysis}\label{ngram}

We analyzed n-grams to extract important insights in text transcriptions to understand language use within news coverage narratives. We extracted bi-grams from text transcriptions of financial market coverage by leveraging the vectors based on the term frequency-inverse document frequency (\textit{tf{-}idf}) technique \cite{Ramos:2003,Gebre:2013}. Specifically, we utilized \textit{tf{-}idf} as a statistical measure to evaluate how important a word is to each text transcription in the corpus; we converted each text transcription into its bag-of-word representation and computed the \textit{tf{-}idf} value of each word using the standard formula, \textit{tf{-}idf}${=}(1+\log{n_{w,t}})\times\log{\frac{T}{T_w}}$, where the \textit{tf{-}idf} value of word $w$ in text transcription $t$ is the log normalization of the number of times the word occurs in the text transcription ($n_{w,t}$) times the inverse log of the number of text transcriptions $T$ and $T_w$ the number of text transcriptions containing word $w$. 

\begin{figure*}[t!]
    \centering
    \includegraphics[scale=0.9]{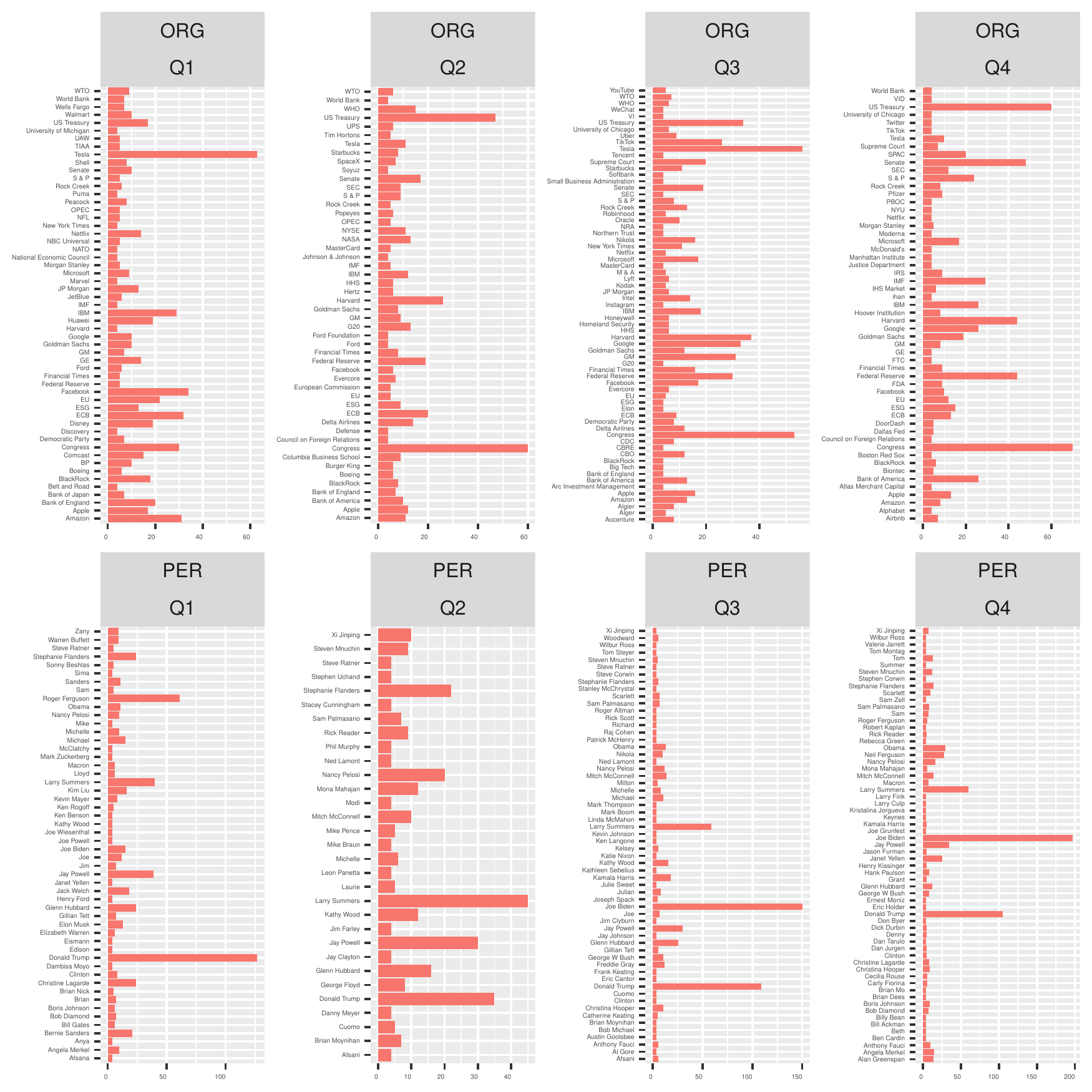}
    \caption{NER results of BLW for 2020}
    \label{fig:ner}
\end{figure*}

\subsection{Topic modeling}\label{topicmodels}

Topic modeling refers to the machine learning task of automatically discovering the abstract `topics' that occur in a collection of documents, and one popular topic modeling technique is known as latent Dirichlet allocation (LDA) \cite{blei:03}. LDA is a probabilistic model that identifies latent topics in a document and can be trained using collapsed Gibbs sampling. Fundamentally, LDA assumes $k$ underlying topics, each of which is a distribution over a fixed vocabulary. While LDA models topics from text corpora \cite{Angelov:20}, it basically suffers from several shortcomings, including difficulty in setting the parameter k (which refers to the number of topics to produce semantically meaningful results), a deficiency in handling short texts \cite{Banda:21}, in capturing the contextual meaning of sentences \cite{ZukZuk:2020}, and its inability to model topic correlations and the evolution of topics over time \cite{WangMcCallum:06}.

\begin{figure*}[t!]

  \begin{subfigure}[b]{0.5\linewidth}
    \centering
    \includegraphics[scale=0.192]{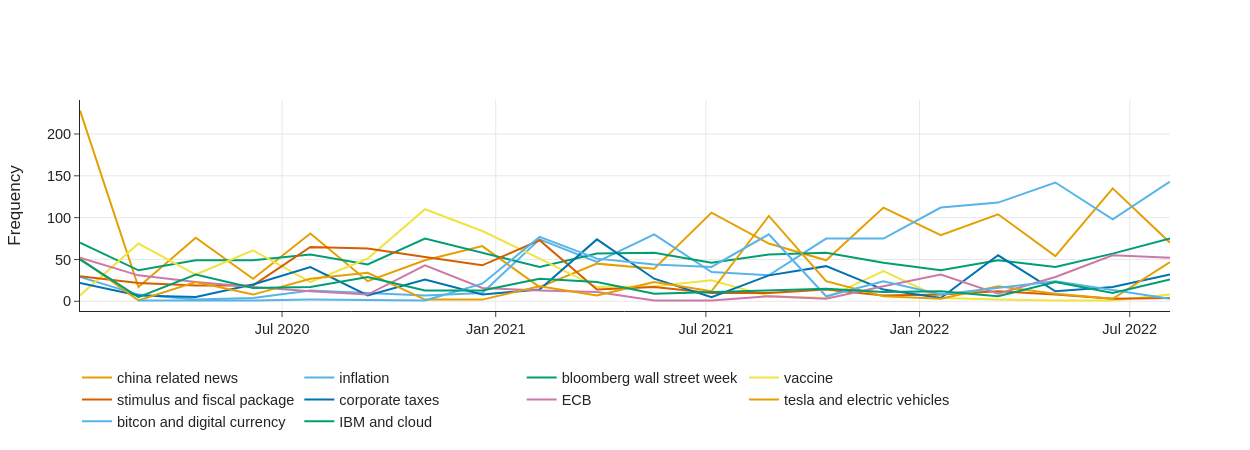} 
    \caption{Topics ranging from 1 to 10} 
    \label{blw:a} 
  \end{subfigure}
  \begin{subfigure}[b]{0.5\linewidth}
    \centering
    \includegraphics[scale=0.192]{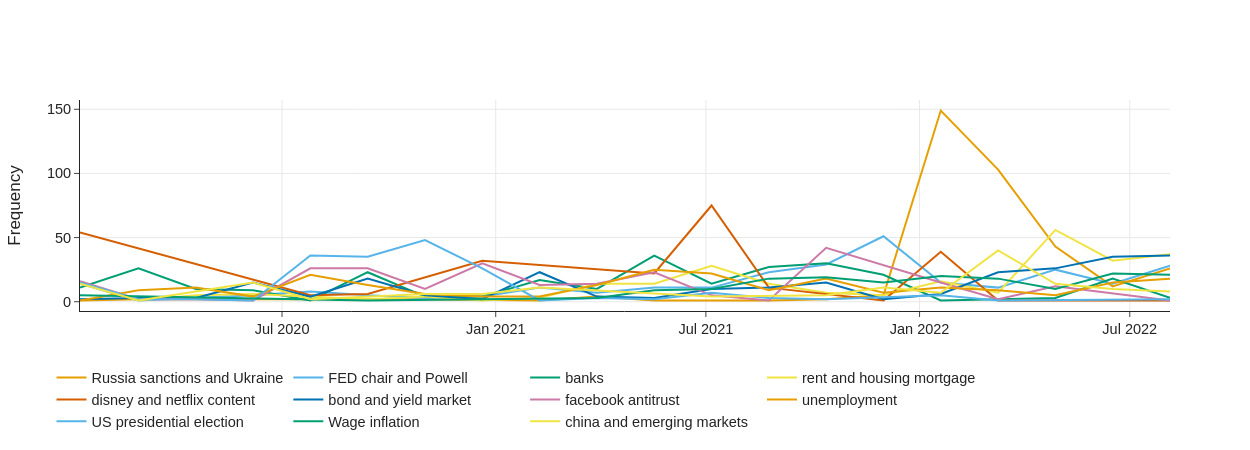} 
    \caption{Topics ranging from 11 to 20}
    \label{blw:b} 
  \end{subfigure} 
  
    \begin{subfigure}[b]{0.5\linewidth}
    \centering
    \includegraphics[scale=0.192]{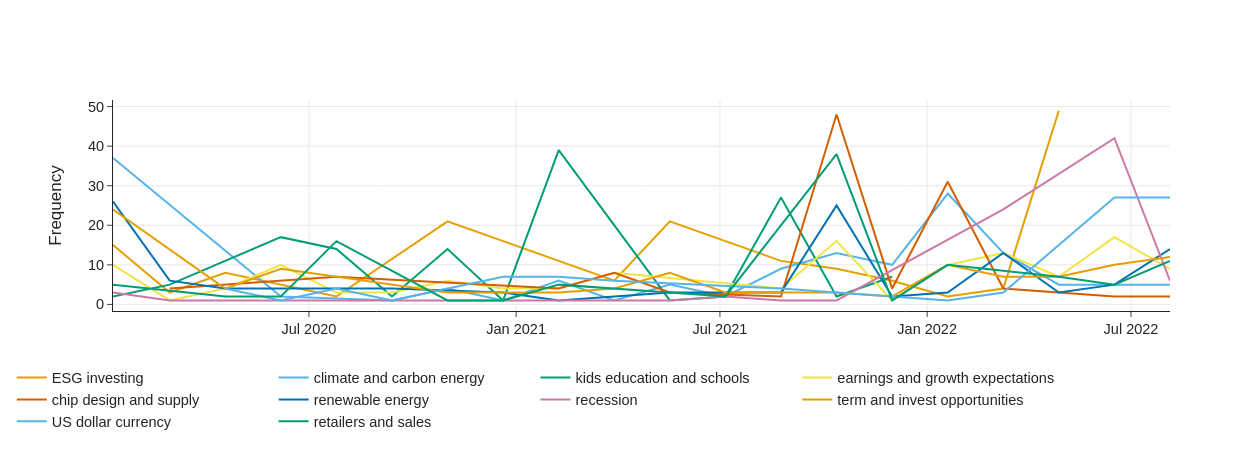} 
    \caption{Topics ranging from 21 to 30} 
    \label{blw:c} 
  \end{subfigure}
  \begin{subfigure}[b]{0.5\linewidth}
    \centering
    \includegraphics[scale=0.192]{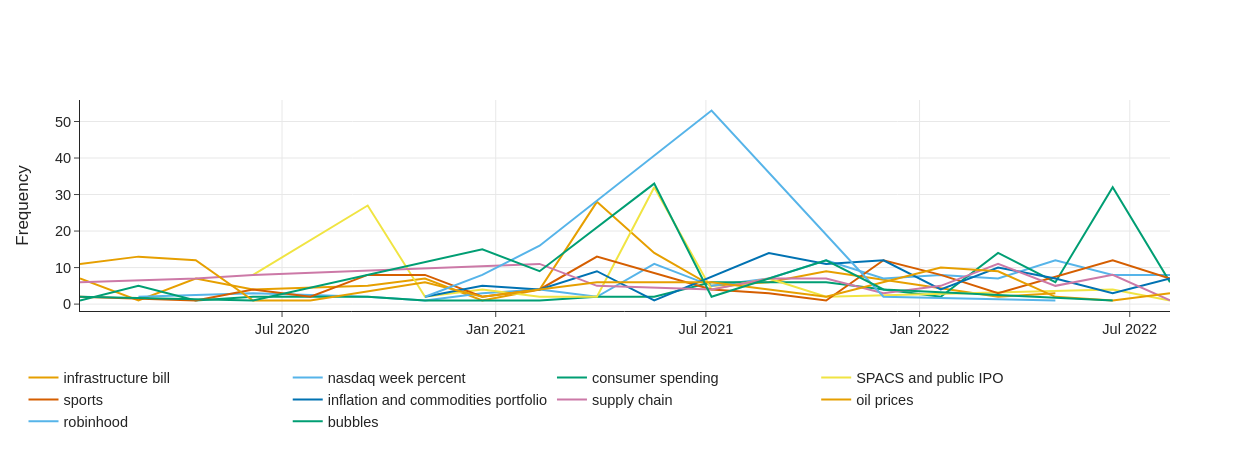} 
    \caption{Topics ranging from 31 to 40}
    \label{blw:d} 
  \end{subfigure} 
  
    \begin{subfigure}[b]{0.5\linewidth}
    \centering
    \includegraphics[scale=0.192]{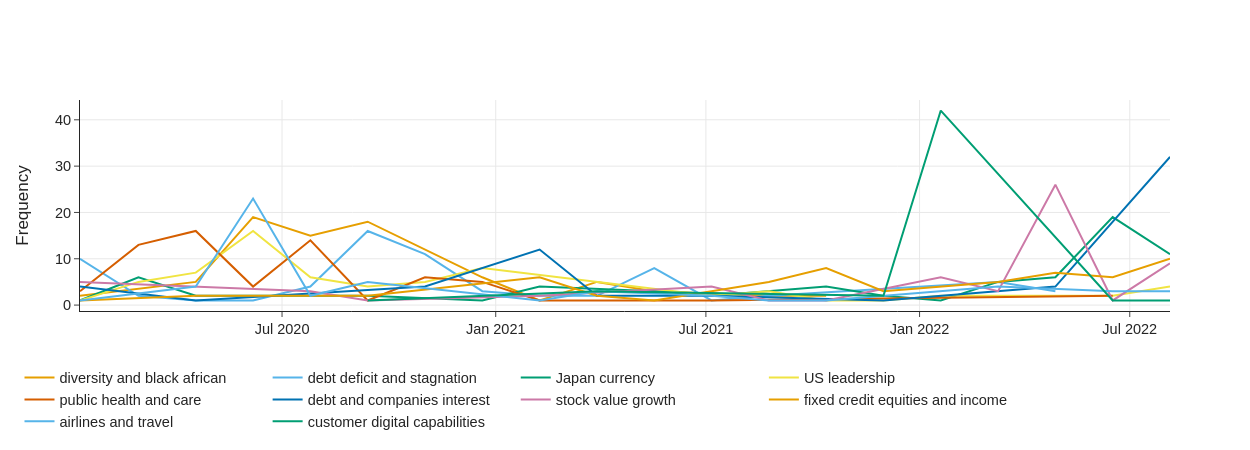} 
    \caption{Topics ranging from 41 to 50} 
    \label{blw:e} 
  \end{subfigure}
  \begin{subfigure}[b]{0.5\linewidth}
    \centering
    \includegraphics[scale=0.192]{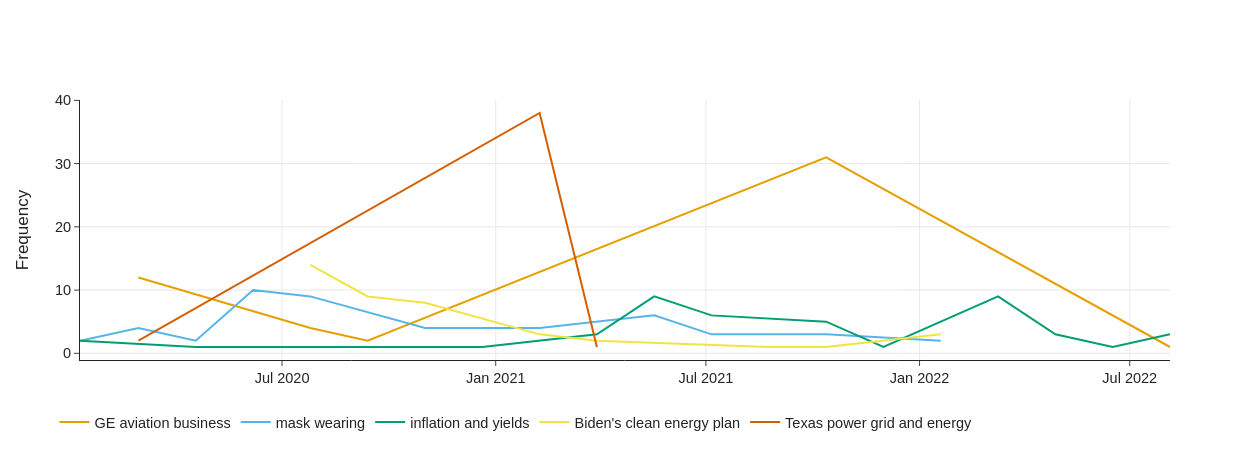} 
    \caption{Topics ranging from 51 to 55}
    \label{blw:f} 
  \end{subfigure} 

\caption{Top 55 topics with frequencies over time extracted from Bloomberg Wall Street Week}\label{figure_blw}  
\vspace{-5mm}
\end{figure*}

To overcome these limitations, the new generation of topic models \cite{Peinelt:2020,Bianchi:2020,Angelov:20,Grootendorst:22} utilize pre-trained representations such as BERT to enable topic modeling (i) to consider the contextual meaning of sentences for supporting the results in order to match the adequate topics and (ii) to include more features for efficiently modeling topic correlations and topic evolution over time. Recent pre-trained contextualized representations like BERT have pushed the state-of-the-art in several areas of natural language processing due to their ability to expressively represent complex semantic relationships from being trained on massive datasets. BERT is a bidirectional Transformer-based pre-trained contextual representation using masked language modeling objective and next sentence prediction tasks \cite{Devlin:2019}. The significant advantage of BERT is that it simultaneously gains the context of words from both the left and right context in all layers. To this end, BERT utilizes a multi-layer bidirectional Transformer encoder, where each layer contains multiple attention heads. 

In this paper we use BERTopic \cite{Grootendorst:22} to generate topics addressed in financial market coverage, analyze the evolution of these topics over time and discover similarities between the topics addressed in BLW, BSM and YFM (\textit{RQ1}). BERTopic leverages BERT embeddings and a class-based term frequency-inverse document frequency to create dense clusters to detect unique topics. In addition, BERTopic generates the topic representations at each timestamp for each topic. The traditional LDA model requires a predefined $k$ (the number of topics) for algorithms to cluster corpus around $k$ topics \cite{blei:03}. BERTopic does not require a predefined $k$, reducing the need for various iterations of model finetuning. The performance of BERTopic over LDA-like models and other topic modeling techniques is reported in \cite{Grootendorst:22}.\footnote{The Python package of BERTopic:\\ \url{https://github.com/MaartenGr/BERTopic}}

\begin{figure*}

  \begin{subfigure}[b]{0.5\linewidth}
    \centering
    \includegraphics[scale=0.192]{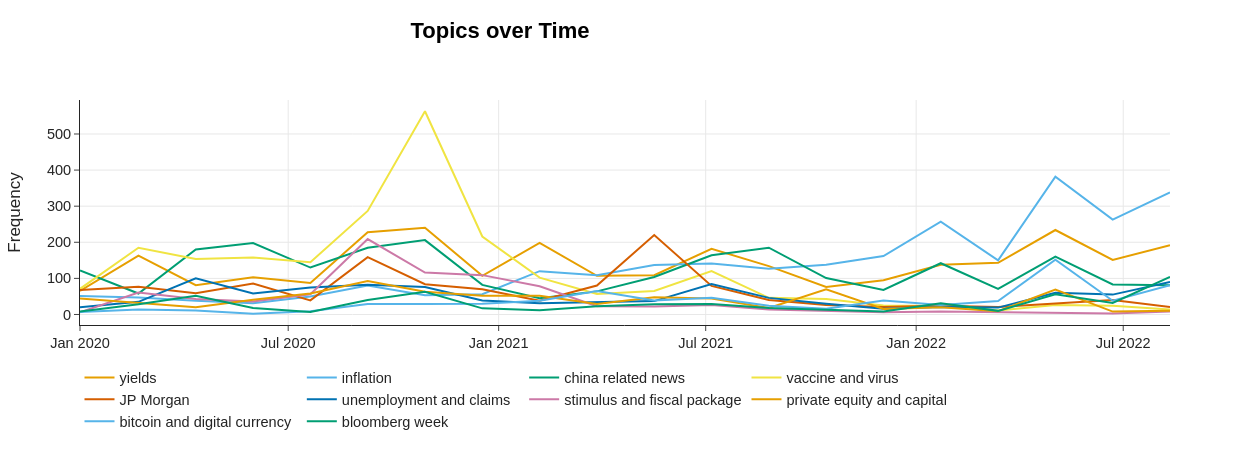} 
    \caption{Topics ranging from 1 to 10}
    \label{bsm:a} 
  \end{subfigure}
  \begin{subfigure}[b]{0.5\linewidth}
    \centering
    \includegraphics[scale=0.192]{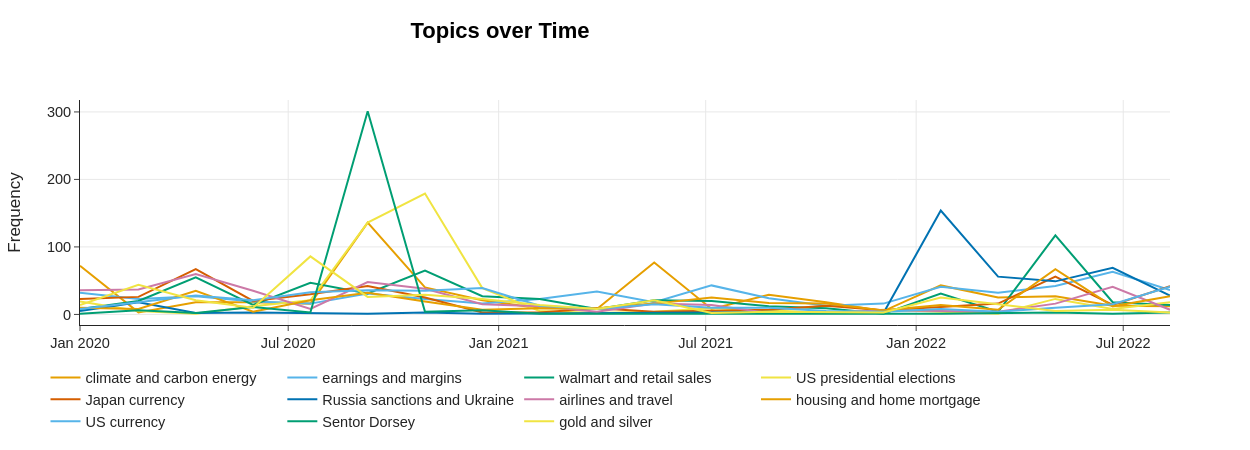} 
    \caption{Topics ranging from 11 to 20}
    \label{bsm:b} 
  \end{subfigure} 
  
    \begin{subfigure}[b]{0.5\linewidth}
    \centering
    \includegraphics[scale=0.192]{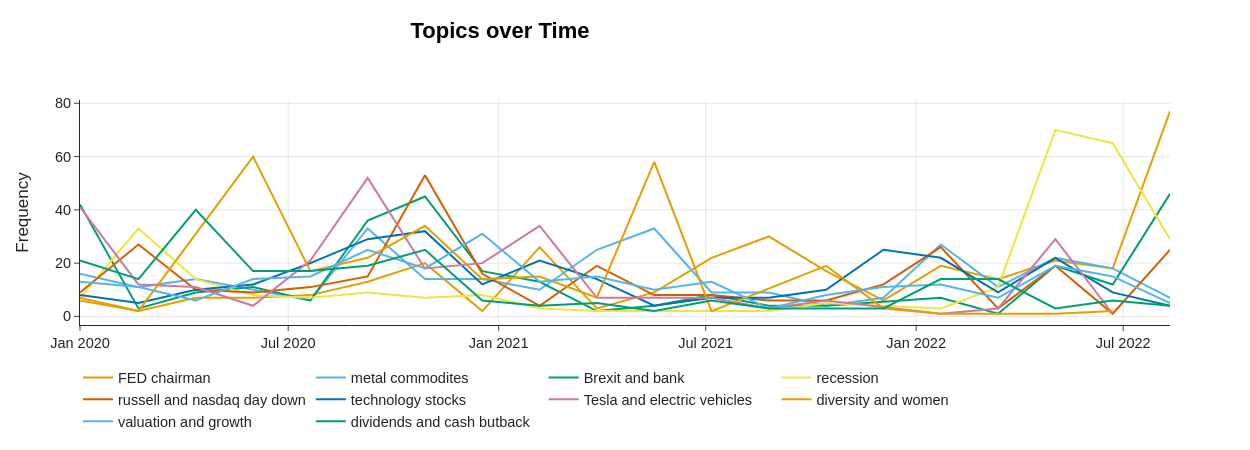} 
    \caption{Topics ranging from 21 to 30} 
    \label{bsm:c} 
  \end{subfigure}
  \begin{subfigure}[b]{0.5\linewidth}
    \centering
    \includegraphics[scale=0.192]{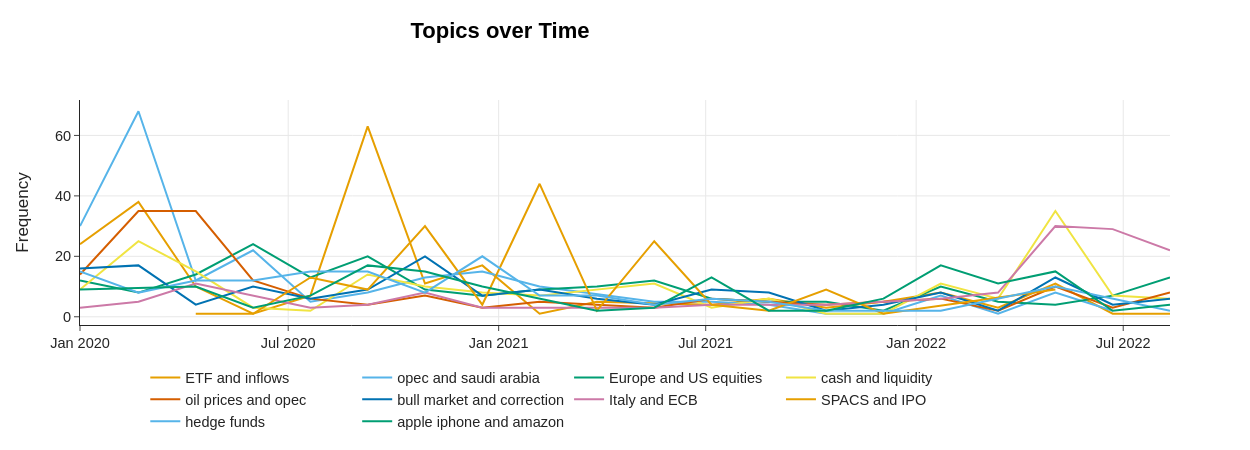} 
    \caption{Topics ranging from 31 to 40}
    \label{bsm:d} 
  \end{subfigure} 
  
    \begin{subfigure}[b]{0.5\linewidth}
    \centering
    \includegraphics[scale=0.192]{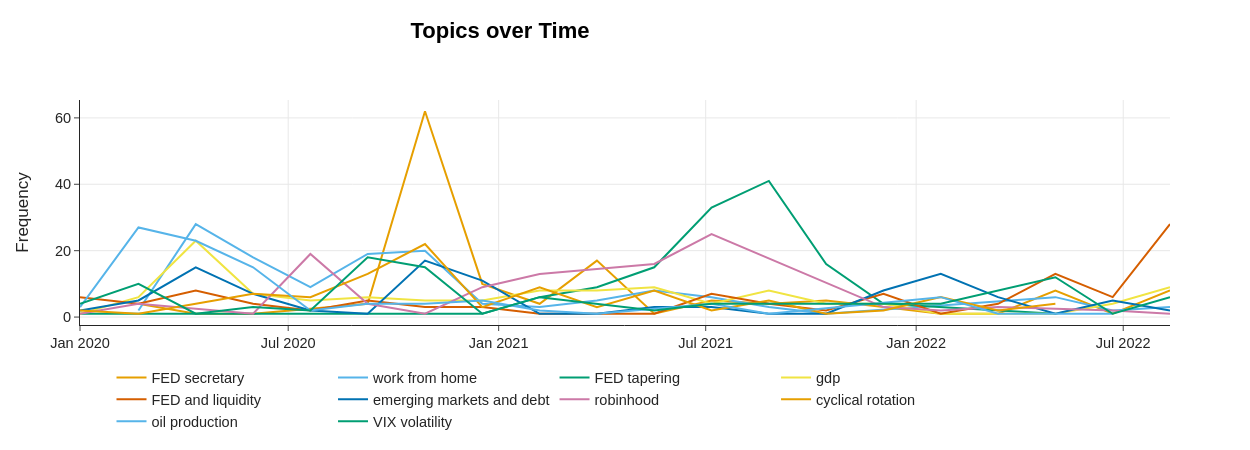} 
    \caption{Topics ranging from 41 to 50} 
    \label{bsm:e} 
  \end{subfigure}
  \begin{subfigure}[b]{0.5\linewidth}
    \centering
    \includegraphics[scale=0.192]{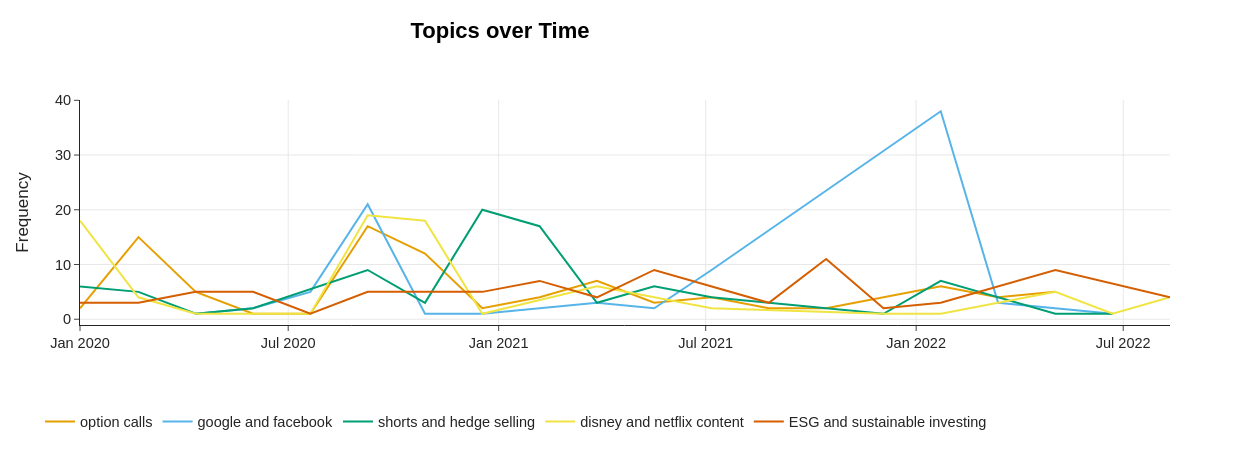} 
    \caption{Topics ranging from 51 to 55}
    \label{bsm:f} 
  \end{subfigure} 

\caption{Top 55 topics with frequencies over time extracted from Bloomberg Stock Market News and Analysis}\label{figure_bsm} 
\end{figure*}

\begin{figure*}

  \begin{subfigure}[b]{0.5\linewidth}
    \centering
    \includegraphics[scale=0.192]{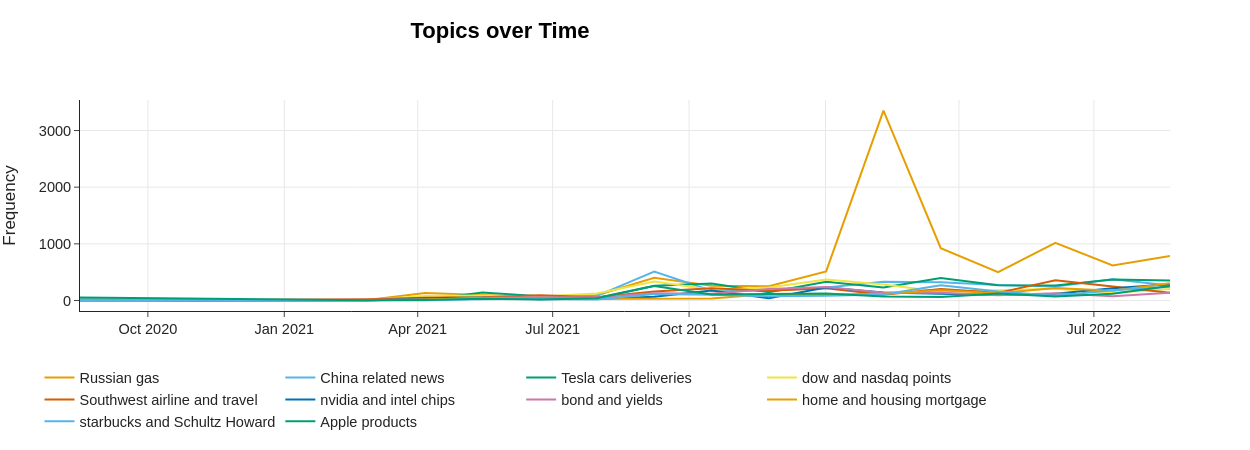} 
    \caption{Topics ranging from 1 to 10} 
    \label{yh:a} 
  \end{subfigure}
  \begin{subfigure}[b]{0.5\linewidth}
    \centering
    \includegraphics[scale=0.192]{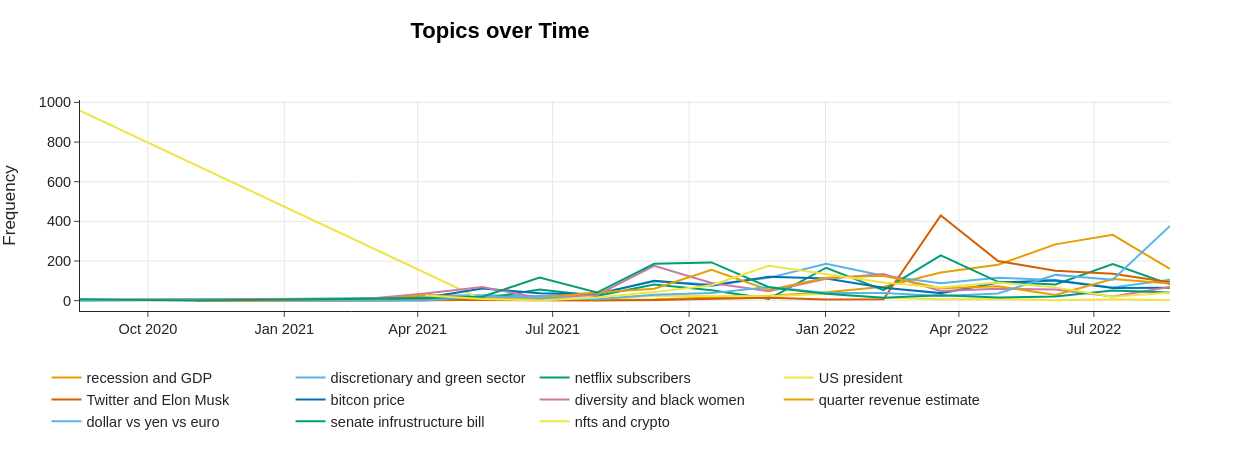}  
    \caption{Topics ranging from 11 to 20}
    \label{yh:b} 
  \end{subfigure} 
  
    \begin{subfigure}[b]{0.5\linewidth}
    \centering
    \includegraphics[scale=0.192]{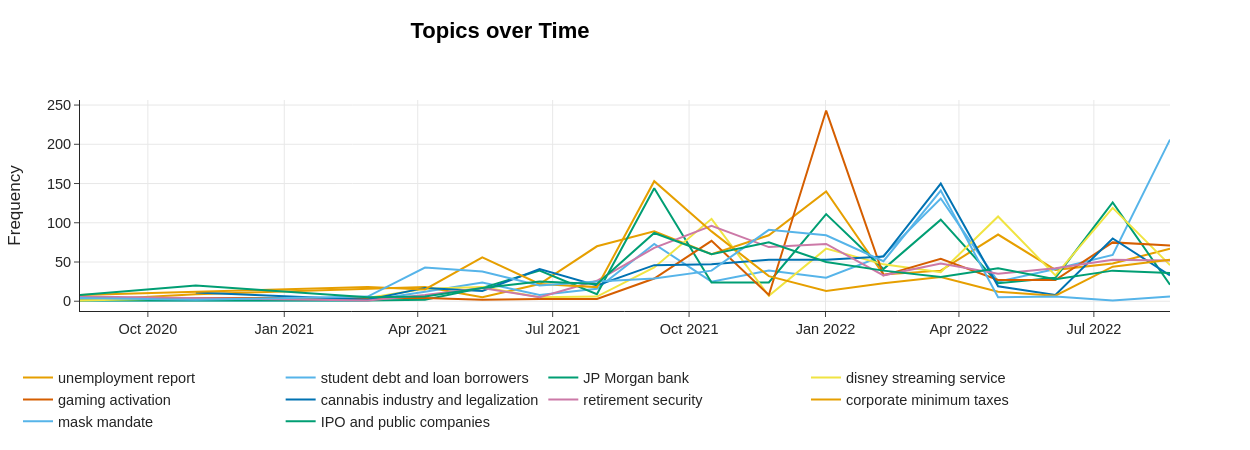} 
    \caption{Topics ranging from 21 to 30} 
    \label{yh:c} 
  \end{subfigure}
  \begin{subfigure}[b]{0.5\linewidth}
    \centering
    \includegraphics[scale=0.192]{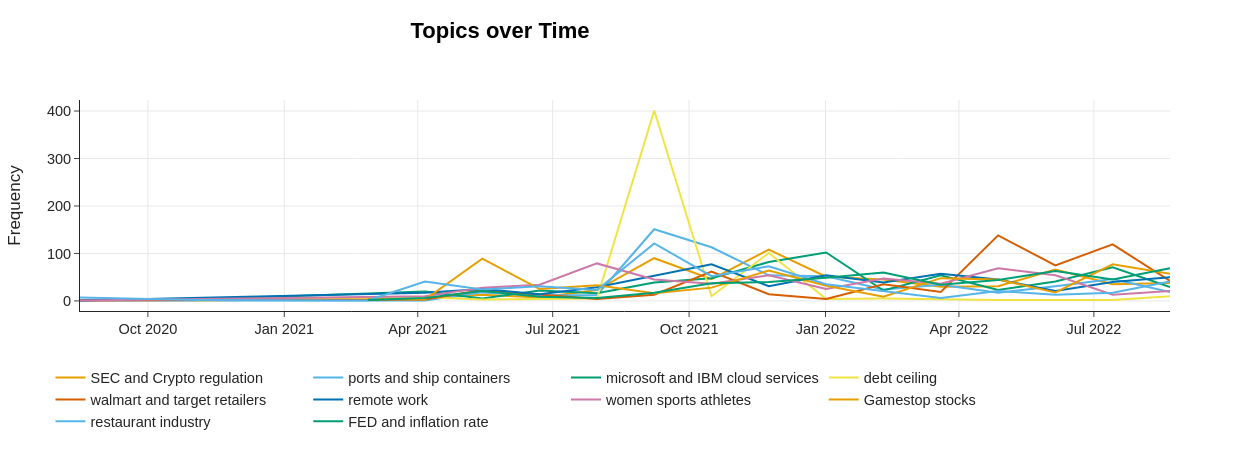} 
    \caption{Topics ranging from 31 to 40}
    \label{yh:d} 
  \end{subfigure} 
  
    \begin{subfigure}[b]{0.5\linewidth}
    \centering
    \includegraphics[scale=0.192]{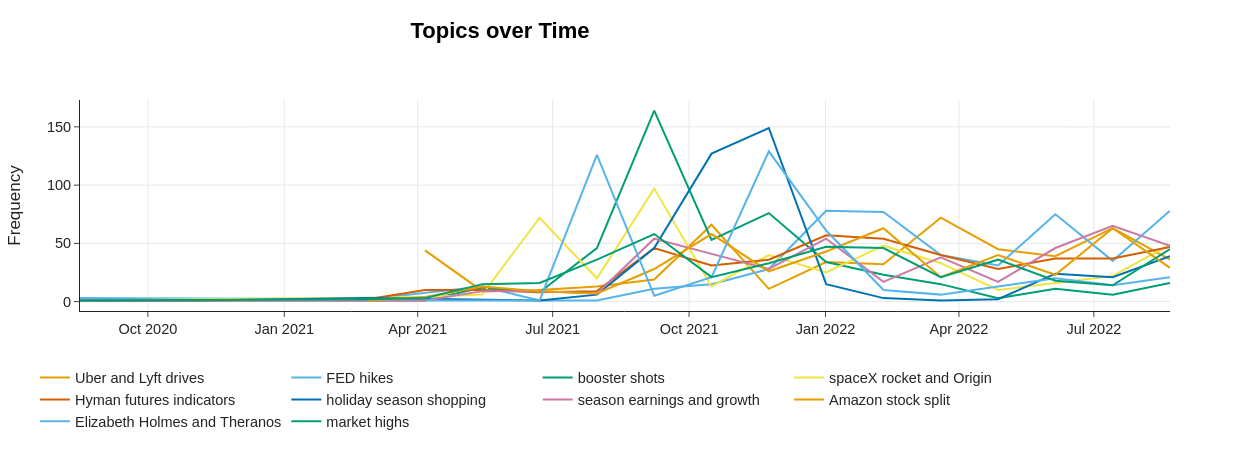} 
    \caption{Topics ranging from 41 to 50}
    \label{yh:e} 
  \end{subfigure}
  \begin{subfigure}[b]{0.5\linewidth}
    \centering
    \includegraphics[scale=0.192]{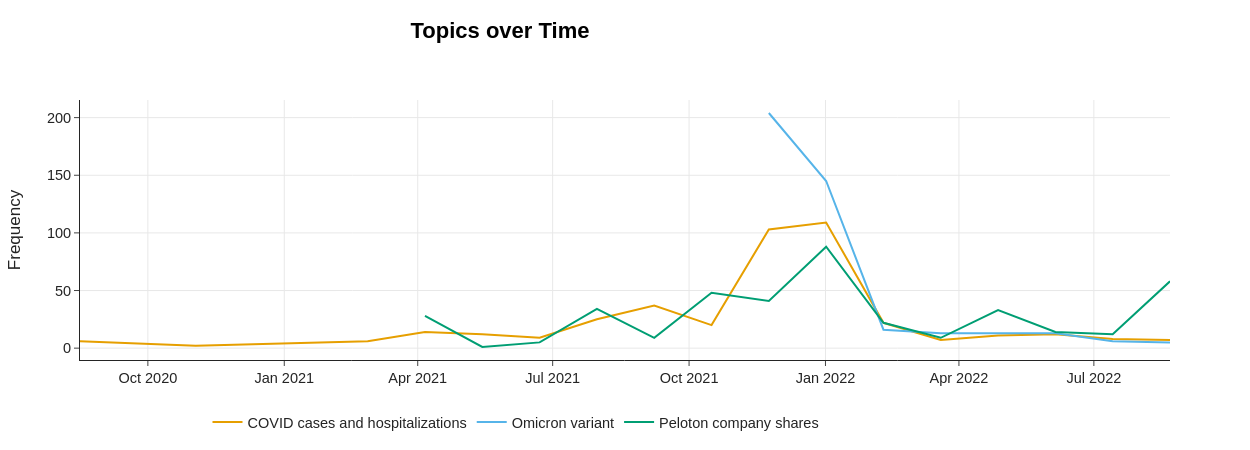} 
    \caption{Topics ranging from 51 to 53}
    \label{yh:f} 
  \end{subfigure} 

\caption{Top 55 topics with frequencies over time extracted from Yahoo Finance Stock Market Coverage}\label{figure_yfm} 
\end{figure*}

\subsection{Named entity recognition}\label{ner}

Named entity recognition (NER) aims at finding and categorizing specific entities in text with their corresponding semantic types such as person names, organizations (such as companies, government organizations, etc.), locations (such as cities, countries, etc.), or date
and time expressions \cite{Li:2020,Perera:2020}. In this paper, we utilized NER to extract the names of persons and organizations mentioned in financial market coverage, and map the frequency of the most mentioned entities over time (\textit{RQ2}). The rationale behind the extraction of NER entities is to identify entities that constitute the center of attention in the financial market and dominate the financial world at a specific time-step. This could support the understanding of the evolution of the topics addressed over time and indicate the entities around which the topics are concentrated in. Note that we removed some names of Bloomberg and Yahoo anchors that appeared in the NER results.

We used a fine-tuned BERT model called bert-base-NER.\footnote{Find the official page of bert-base-NER on HuggingFace, dslim/bert-base-NER} In this paper we used the initial model of bert-case-NER without modifying its architecture or implementation. It is important to note that bert-base-NER is ready to use for NER and achieves state-of-the-art performance for the NER task, and it has been trained to extract four types of entities: location (LOC), organizations (ORG), person (PER) and miscellaneous (MISC). Specifically, bert-base-NER is a version of the bert-base-cased model that was fine-tuned on the English version of the standard CoNLL-2003 Named Entity Recognition dataset \cite{Sang:2003,Devlin:2019}.

\section{Results}

This section first investigates the relationships between words using n-grams (n=2) and language use within news coverage narratives and provides sentiments of n-grams indicating economic and financial concerns. Second, topic modeling and name entity recognition are performed to examine the evolution of topics over time and discern the coverage narratives, and then understand the implication of the most mentioned persons and organizations in the stock market.

\noindent
\textbf{Bi-gram analysis.}
As mentioned in \S\ref{ngram}, we sort to determine the link between words by looking at which words typically come after others right away. By utilizing the \textit{tf-idf} statistical measure, we identify the importance of each consecutive sequence of words in each year's corpus. Figures \ref{fig:coverage_bws_ngrams}, \ref{fig:coverage_bsm_ngrams} and \ref{fig:coverage_yfm_ngrams} show the top 55 bi-grams drawn from the market coverage. Generally, each year, we observe the bi-grams are primarily on economic and financial markets-related topics as well as some pertaining issues that happened during the years.

While examining the results obtained from the BLW, BSM and YFM datasets, we observe that the majority of bi-grams indicate economics-related topics except for \textit{``divided government'', ``delta variant'', ``democratic national''} and \textit{``blue wave''}, among others. Particularly, we note that these datasets assimilate topics/events pertaining to aspects of finance that have major impacts on the economy, for example, prices. Besides these, other interesting financial discourses are centered on the \textit{“game stop"} fiasco in 2021 \cite{Malz:21};
we observed this as a bi-gram in BSM.
{\color{black} Interestingly, we note the presence of the bi-gram \textit{``digital assets''} along with \textit{``cyber security''}. Recent turmoil in the cryptocurrency market has underlined the critical risks involved with investing in or engaging with digital assets. Digital assets raise cybersecurity concerns requiring regulatory controls and measures to protect individuals from cybercrime and other critical risks \cite{Bauer:20}. We observe an important number of cryptocurrency-related results, including bi-grams \textit{``bitcoin futures''} and \textit{``crypto space''} in 2021, topics \textit{``bitcoin price''} and \textit{``nfts and crypto''} in Figure \ref{figure_yfm} and the topic \textit{``bitcoin and digital currency''} in Figures \ref{figure_blw} and \ref{figure_bsm}.}

We report bi-grams highlighting recent events occurring in Ukraine as well as their continuous in the early months; these bi-grams include \textit{“russia ukraine", “russia oil"} and \textit{“ukraine conflict"} are especially inline with the commodities market.\footnote{\href{https://shorturl.at/lqWZ3}{shorturl.at/lqWZ3 Accessed 23 December 2022}}
Additionally, we note that the coverage bi-grams also identify persons and organizations including \textit{``central banks'', ``paul krugman''} and \textit{``gary gensler''}. For instance, Paul Krugman is an economist and a contributor on Bloomberg\footnote{\href{https://shorturl.at/dHIX8}{shorturl.at/dHIX8 Accessed 23 December 2022}} and Gary Gensler Chairperson of the U.S. Securities and Exchange Commission since 2021.\footnote{https://www.sec.gov/} 
Overall, we find that some bi-grams depict the language use in coverage which attributes to events such as \textit{``president elect'' or ``rate hikes''}; each referring to the general elections and (imminent) announcement of interest rate hikes and discussions on these type of events.\footnote{\href{https://shorturl.at/mwBGL}{shorturl.at/mwBGL Accessed 23 December 2022}}

Figure \ref{fig:sentiment_bws}, \ref{fig:sentiment_bsm} and \ref{fig:sentiment_yfm} show the top 6 economic (financial)-related keywords identified in the bi-gram that best describes the financial markets. The keywords (\textit{consumer, economy, inflation, loan, market} and \textit{recession}) are not exhaustive and can be expanded. The choice of these keywords is arbitrary, we believe that they reflect and pertain to broad discussions regarding recent events. We examine how frequently sentiment-associated words are preceded by these keywords, which attribute to positive or negative sentiments; {\color{black}with positive or negative values indicating the direction of the sentiment}.  
{\color{black}We note that bi-grams stemming from the previously mentioned keywords identify the most common economic events}. For instance, the bi-gram of \textit{``demand consumer''} has a negative sentiment while \textit{``confidence consumer''} has a positive sentiment. These bi-grams reflect the events of supply-chain issues or consumers' ability to buy items.\footnote{\href{https://shorturl.at/brJOP}{shorturl.at/brJOP Accessed 23 December 2022}} 
Further, \textit{``inflation''} and \textit{``economy''} discourse cite either \textit{``growth/growing/good''} or \textit{``risk(s)/worse/stalling''} painting the picture of positive and negative sentiments, respectively. The bi-gram of \textit{``stalling economy''} essentially describes one with a growth rate below some threshold level. Thus, the possible effects of the COVID-19 pandemic. The bi-gram of \textit{``growth loan''} is the maximum positive sentiment across BLW, BSM and YFM data. Our analysis indicates the right direction of such a bi-gram. However, the financial keyword, \textit{``market's''} bi-gram has \textit{“share"} as the most positive sentiment and with \textit{``demand/debt''} as the opposite. A discussion relating to market share could be attributed to an organization as the bi-grams of Figures \ref{fig:coverage_bws_ngrams},\ref{fig:coverage_bsm_ngrams} and \ref{fig:coverage_yfm_ngrams} identify some organizations. Interestingly, we observe that the NER analysis also identified such organizations.

\noindent
\textbf{Named entity recognition.} 
Within the context of (financial) news coverage, individuals (or persons) are either introduced as panelists or as a contributor or mentioned (cited) to affirm a statement. Likewise, some individuals are associated with some organization (or institution), or sometimes discussions are centered around an organization based on what might be trending. To distinguish between persons and organizations from our corpora, we employed a NER model as described in \S\ref{ner}. Figure \ref{fig:ner} shows the distinct entities for the BLW corpus in each quarter of 2020.

Each quarter shows the frequencies of the top 60 entities (organizations and persons). A closer look at all the quarters' organizations reveals the following organizations having the highest mentions or often discussed: \textit{Tesla, Congress} and \textit{US Treasury}. Note that \textit{Congress} represents both the ``House of Representatives" and ``Senate". Within the various quarters, we noticed that some major technology companies were frequently in discussions: \textit{IBM, Huawei, Facebook, Apple} and \textit{Amazon}. The \textit{Congress}, for example, were often concerning stimulus package discussions.\footnote{www.bloomberg.com/news/articles/2020-03-25/what-s-in-congress-2-trillion-coronavirus-stimulus-package} We further observed that the identified organizational entities are either governmental or private financial institutions, such as \textit{US Treasury, Goldman Sachs} and \textit{Rock Creek}. Besides, there were organizations of global and continental significance during the coverage. For example, the \textit{World Bank, OPEC} and \textit{ECB}---the European Central Bank.

Similarly, the news coverage on persons ranges from world leaders or politicians to investment moguls to financial experts to heads of institutions and others. In the first quarter of 2020, we noticed that the name of \textit{Donald Trump}, the then president of the United States of America (USA), was frequently mentioned; this suggests that the events of January 6, 2020, and subsequent events had tremendous discussions in the financial and economic news space. We also noticed an important frequency around the name of \textit{Roger Ferguson}, the former president and Chief Executive Officer (CEO) of \textit{TIAA}---organization and a contributor on BLW. In the subsequent quarters, \textit{Larry Summers}, a renowned financial expert and contributor; \textit{Joe Biden}, the current president of the USA; and \textit{Donald Trump} were highly mentioned. Concerning \textit{Larry Summers}, he provides insight into how prospective the economic and financial outlook would be based on some announcements. Of particular notice was \textit{George Floyd} in the financial news coverage in the second quarter of 2020; his murder sparked numerous protests and moments of reckoning that reverberated far beyond the United States. Based on the NER sample result, we observed that financial news coverage does not only cover finance and economic topics but also general topics. In the next section, we identify some common topics from the news coverage. 

\noindent
\textbf{Topic modeling.}
Figures \ref{figure_blw}, \ref{figure_bsm} and \ref{figure_yfm} show the top 55 salient topics from the BLW, BSM and YFM, respectively. Figures are organized into six sub-figures at the rate of ten topics per sub-figure to provide a better visualization of the frequency of topics over time for the period from January 2020 to September 2022. 

Figure \ref{blw:a} shows the ten most topics addressed in BLW. These topics include ``\textit{inflation}'', ``\textit{vaccine}'', ``\textit{China-related news}'' and ``\textit{Tesla and electric vehicles}. Particularly, for the topic ``\textit{Tesla and electric vehicles}'', a high spike was observed in early 2020, followed by a drop in frequency over the first quarter of 2021. Even though electric vehicles seemed less frequently discussed in favor of topics related to the vaccine and COVID-19, we observed that the discussions around electric vehicles remained one of the most salient topics in the market coverage. We noticed many spikes in Figure \ref{blw:a} and Figure \ref{bsm:a} for the topic ``\textit{vaccine}'' during 2020 and the topic ``\textit{mask-wearing}'' in Figure \ref{yh:c}. One of the reasons that could partly justify this observation is that pharmaceutical companies such as Pfizer and BioNTech started research on developing vaccines for COVID-19 during that period and announced promising results. In November 2020, Pfizer announced the vaccine releases, followed by a vaccination campaign worldwide.\footnote{\href{https://www.pfizer.com/news/press-release/press-release-detail/pfizer-and-biontech-announce-vaccine-candidate-against}{Pfizer and BioNTech Announce Vaccine Candidate Against COVID-19. \url{shorturl.at/mHNV4}}} The COVID-19 pandemic has caused major social and economic impacts on the lives of people across the world. One of the direct impacts includes unemployment in the labor market~(Figure \ref{blw:b}, \ref{bsm:a}, and \ref{yh:c}), and inflation, in the financial market. Figure \ref{yh:c} shows an increase in frequency over time for the topic ``\textit{inflation}'' from 2021 to 2022. We note that this topic received considerable attention in the market coverage along with its related topics, such as ``\textit{home and housing mortgage}'', due to the increase in mortgage rates, and ``\textit{recession}''.

Figures \ref{blw:b}, \ref{bsm:a} and \ref{yh:c} highlight the evolution of topics over time for topics such as ``\textit{Disney and Netflix content}'' and ``\textit{Russia sanctions and Ukraine}''. The Russian invasion of Ukraine in early 2022 caused knock-on effects worldwide. Sanctions imposed on Russia by the United States and other countries engendered multilateral effects on the global economy in general and the stock market in particular. This reason could be retained as a compelling justification to partly explain the many high spikes that we observed for the topic of Russia and Ukraine. The topic ``\textit{Disney and Netflix content}'' indicates its large surge during the period of the first COVID-19 lockdown. Note that lockdown was one of the restrictive measures taken by governments to contain the ongoing pandemic. During the lockdown period, many people spent most of their time on streaming platforms as they could not go out. Online streaming platform subscriptions have increased along with their corresponding stock price. 

\section{Discussion}
News coverage provides contest and analysis needed to aid viewers in ascertaining further insight lacking from other news sources (newspapers or blogs) through anchors and guests (experts') discussions. In this paper, we collected news coverage data from YouTube and Bloomberg related to financial and economic news to identify the most discussed topics from transcribed video news coverage.

The primary goal of our research was to identify the similarities of news coverage topics regarding major financial events across different news channels. Our findings describe the usefulness of considering video (visual) as a data source. By analyzing the similarity between other channels, we observe some related bi-gram keywords and entities (organizations and persons). The bi-gram provides an overview of the structure of language use during news coverage through discussions and headlines briefing of news segments. Our results find that news coverage evolved, and discussions were often centered more on recent events surrounding specific financial markets.

Secondly, we identified major financial events through the evolution of topics over time and their frequencies. Our topic models broadly reflect the evolution and variation of topics related to financial events. Important to note are the global events documented in various studies that are in tandem with financial markets, such as the Russo-Ukrainian War \cite{Lo:22}. 
Prior work found the effect of news coverage on trading and prices \cite{Engelberg:11,Haroon:20}, while our results identify the narrative of news coverage without any relation to either trading behavior or price volatility. Our results can be used to create dashboards portraying outputs stemming from financial market coverage from various reliable media channels. This can help anticipate "investment" actions or predict market pricing based on the news coverage and identify the most frequently cited entities to make a good investment choice. Further, the results investigated the less frequently cited entities, which one can keep a constant eye on or keep on track to ensure if they constitute a new market opportunity and if something might skyrocket overnight.

\section{Conclusion}
In this paper, we characterize financial market coverage from YouTube. To this end, we utilize OpenAI's Whisper speech-to-text model to generate a text corpus of market coverage YouTube videos from Bloomberg and Yahoo Finance. Then, we use natural language processing to gain insights into language usage in financial market coverage. Additionally, we investigate the prevalence and evolution of trending topics and the influence of certain persons and organizations on the financial market. We discover similarities between topics and exhibit content coordination regarding major financial events. Through this characterization, we gain a better understanding of the dynamics of financial market coverage and valuable insights into current financial events and the global economy. We show how our findings can be used to predict market performance and pricing and to support investment actions and decision-making. In the future, we would like to experiment with market forecasts using a holistic model that combines financial market coverage and stock prices and includes features such as n-gram, NER, topic modeling, and emotions.



\bibliography{custom}
\bibliographystyle{acl_natbib}

\end{document}